\documentclass{jpsj-suppl}
\usepackage{times}
\usepackage{color}
\usepackage{textcomp}
\usepackage{epstopdf}

\title{Anomalous Microwave Surface Resistance of CeCu$_6$}

\author{Daniel Hafner$^{1}$, Martin Dressel$^{1}$, Oliver Stockert$^{2}$, Kai Grube$^{3}$, Hilbert v. L\"ohneysen$^{3,4}$, Marc Scheffler$^{1}$\thanks{E-mail address: scheffl@pi1.physik.uni-stuttgart.de}
}
\inst{$^{1}$1. Physikalisches Institut, Universit\"at Stuttgart, D-70550 Stuttgart, Germany\\
$^{2}$Max-Planck-Institut f\"ur Chemische Physik fester Stoffe, D-01187 Dresden, Germany\\
$^{3}$Institut f\"ur Festk\"orperphysik, Karlsruhe Institute of Technology (KIT), D-76131 Karlsruhe, Germany\\
$^{4}$Physikalisches Institut, Karlsruhe Institute of Technology (KIT), D-76131 Karlsruhe, Germany
}

\recdate{October 1, 2013}

\abst{We present surface resistance measurements of the archetypical heavy-fermion compound CeCu$_6$ for frequencies between 3.7 and 18 GHz and temperatures from 1.2 to 6~K. The measurements were performed with superconducting stripline resonators that allow simultaneous measurements at multiple frequencies. The surface resistance of CeCu$_6$ exhibits a pronounced decrease below 3~K, in consistence with dc resistivity. The low-temperature frequency dependence of the surface resistance follows a power law with exponent 2/3. While for conventional metals this would be consistent with the anomalous skin effect, we discuss the present situation of a heavy-fermion metal, where this frequency dependence might instead stem from the influence of electronic correlations.}

\kword{CeCu$_6$, microwave surface resistance, heavy fermions, skin effect}

\begin{document}
\maketitle

\section{Introduction}
The peculiar properties of heavy-fermion metals are due to the interaction of magnetic moments and conduction electrons. Electrical charges can directly be probed by optical spectroscopy in the appropriate frequency range.\cite{Basov2011,Scheffler2013} This can indicate characteristic excitations, such as the mid-infrared feature discussed in the context of the so-called hybridization gap,\cite{Dordevic2001,Okamura2007,Kimura2009,Sichelschmidt2013} as well as the charge dynamics. Here the important frequency scale is the relaxation rate $\Gamma$ that reflects the scattering mechanisms. While for conventional metals this is found at infrared or THz frequencies,\cite{Dressel2006} for heavy fermions it is much lower: moving from a non-interacting metal with effective electron mass $m$ to heavy fermions with enhanced effective mass $m^*$ goes along with a reduction of the relaxation rate: $\Gamma^*/\Gamma = m/m^*$.\cite{Millis1987} With mass enhancements of up to several hundred in heavy fermions, an extremely low relaxation rate at GHz frequencies is expected.\cite{Millis1987,Scheffler2013} But the experimental situation is not very clear: for two uranium-based compounds, UPd$_2$Al$_3$ and UNi$_2$Al$_3$, broadband microwave spectroscopy found relaxation rates of a few GHz.\cite{Scheffler2005c,Scheffler2006,Steinberg2012} This was surprising because these materials have only moderate mass enhancement (around 70) and because these experiments were performed on thin film samples (with higher residual scattering rates than single crystals).\cite{Scheffler2010} For all other heavy-fermion materials, the experimental situation is much less resolved, as only CePd$_3$, CeAl$_3$, U$_2$Zn$_{17}$, URu$_2$Si$_2$, and UPt$_3$ have been studied at GHz frequencies.\cite{Webb1986,Beyermann1988a,Awasthi1993,Degiorgi1997,Tran2002} The latter were probed with cavity resonators, which are usually operated at a single frequency and hardly yield spectral information. Still, these experiments suggested relaxation rates in the range of 100~GHz, and it is an open question whether the extremely low relaxation rates of UPd$_2$Al$_3$ and UNi$_2$Al$_3$ are exceptional or whether similar relaxation rates could be found in other heavy-fermion compounds.
Another interesting question goes beyond the simple relaxation rate of the Drude framework,\cite{Dressel2006} but instead considers $\Gamma(\omega)$ a function of frequency.\cite{Basov2011} For example, Fermi-liquid theory predicts a quadratic frequency dependence of $\Gamma(\omega)$ (similar to the well-known $T^2$ temperature dependence of $\Gamma$ that governs the dc resistivity of many heavy-fermion materials),\cite{Scheffler2013,Gurzhi1959,Maslov2012,Berthod2013} but so far this $\omega^2$ dependence has not been observed unequivocally in any metal.\cite{Maslov2012,Berthod2013}

These open questions motivate microwave studies on CeCu$_6$: it is a well-established heavy-fermion material with very high effective mass,\cite{Stewart1984,Onuki1987,vonLoehneysen1996} i.e., the relaxation rate should be extremely low. Furthermore, at very low temperatures, CeCu$_6$ exhibits clear $T^2$ behavior in resistivity \cite{Ott1985,Amato1985,vLoehneysen1995,vonLoehneysen1996,vLoehneysen1998} and thus could be a model system for Fermi-liquid optics.\cite{Scheffler2013} Finally, CeCu$_6$ can act as a starting point for future low-energy optics studies of CeCu$_{6-x}$Au$_x$, which is one of the best-established quantum-critical heavy-fermion systems\cite{vonLoehneysen1996,vLoehneysen1998,vLoehneysen2007,Gegenwart2008} and thus would be an interesting material for microwave studies that might address optical non-Fermi-liquid behavior \cite{Scheffler2013} or $\omega/T$ scaling.\cite{Schroeder2000} So far, optical measurements on CeCu$_6$ were limited to infrared frequencies, higher than 1~meV $\approx$ 240~GHz.\cite{Marabelli1990}

The intrinsic low-temperature features expected for the optical response of CeCu$_6$ can only be probed in high-quality single crystals,\cite{Scheffler2013} which in turn require a highly sensitive microwave technique based on a resonator. Since we are particularly interested in the frequency-dependent response, we do not employ traditional cavity resonators, but instead one-dimensional resonators based on superconducting striplines that allow for measurements at several resonance frequencies at the same time.\cite{DiIorio1988,Scheffler2012,Hafner2013}

\section{Experiment}
\begin{figure}[tb]
\begin{center}
\includegraphics{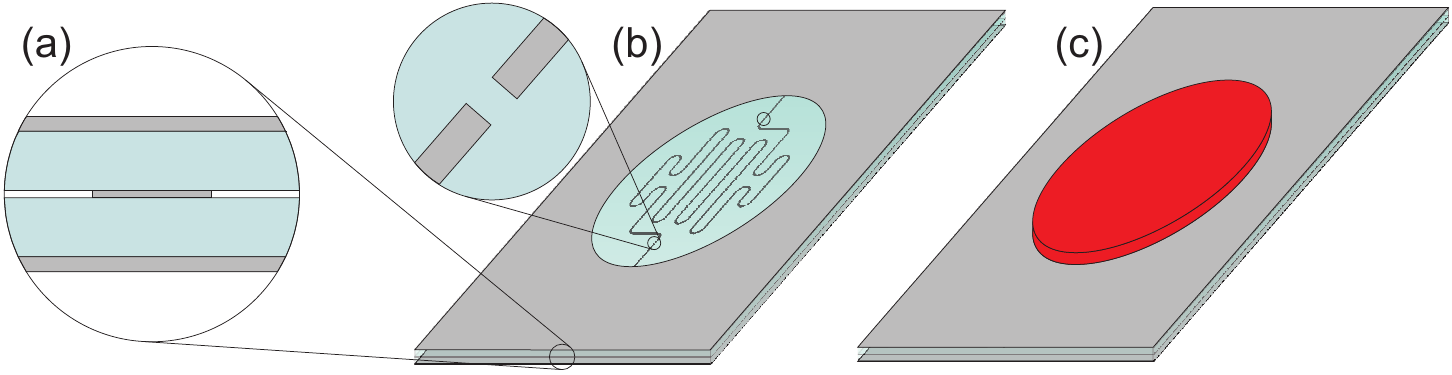}
\end{center}
\caption{(Color online) (a) Schematic cross section of the basic stripline structure. (b) Stripline design with meandered center conductor and two gaps (indicated by circles) to form the resonator. The top ground plane has an opening on top of the resonant section which is eventually covered by the sample as shown in (c).}
\label{f1}
\end{figure}
The stripline is a layered structure, as illustrated in Fig.~\ref{f1}(a): The microwave signal is carried by a center strip sandwiched between two dielectrics and two ground planes.
The dielectric is formed by 127-\textmu m thin 12x10\,mm$^2$ sapphire plates, while 8-\textmu m thin lead foils are used for the ground planes.
The lead center strip is thermally evaporated onto the lower sapphire substrate using laser-cut shadow masks.
Two 50-\textmu m wide bridges on the mask create coupling gaps in the center conductor as seen in Fig.~\ref{f1}(b).
This turns the stripline into a resonant structure with resonance frequencies
\begin{equation}
\nu_{0,n}=\frac{nc}{2\sqrt{\epsilon}L},
\label{eq1}
\end{equation}
where $n$ denotes the mode number, $c$ the speed of light, $\epsilon$ the permittivity of the dielectric and $L$ the length of the stripline section between the gaps.
In order to reach resonance frequencies as low as possible, we meander the center strip as illustrated in Fig.~\ref{f1}(b) for an increased transmission path $L$.
The layers are then stacked into a brass box and connected via stripline launchers to the 50~$\Omega$ measurement circuitry.\cite{Hafner2013}
This setup serves for surface resistance measurements of any conductive bulk sample that is placed as top ground plane onto the resonant section of the stripline as shown in Fig.~\ref{f1}(c). The single-crystalline sample of this study was grown by the Czochralski method under high-purity argon atmosphere and spark-cut into round plates of 6 mm diameter.

To obtain the surface resistance of the sample, we measure the quality factor $Q$ of the resonator.
$Q^{-1}$ is a measure of the overall losses in the resonator. Sapphire as low-loss dielectric \cite{Krupka1999}, superconducting lead for center strip and bottom ground plane (negligible losses well below the critical temperature $T_c= 7.2$~K of lead), and a weak coupling to the outer circuitry ensure that the intrinsic properties of the metallic sample are the dominant loss channel of the resonator.
In this case, the resonator $Q$ is connected to the surface resistance $R_\textrm{s}$ of the sample by a purely geometrical factor $G$ and the resonance frequency $\nu_0$:\cite{DiIorio1988,Oates1991,Hafner2013}
\begin{equation}
R_\textrm{s}=G\frac{\nu_0}{Q}.
\label{eq3}
\end{equation}
For the geometries used in this study, $G$ takes the value $2.68\,\Omega$/GHz.
Measurements were performed using resonators with fundamental frequency of $\nu_{0,1} \approx 1.9\,$GHz and a glass cryostat with dedicated microwave insert at temperatures from 1.2 to 7.2\,K.\cite{Scheffler2005a,Steinberg2012}

\section{Results}

\begin{figure}[tb]
\begin{center}
\includegraphics{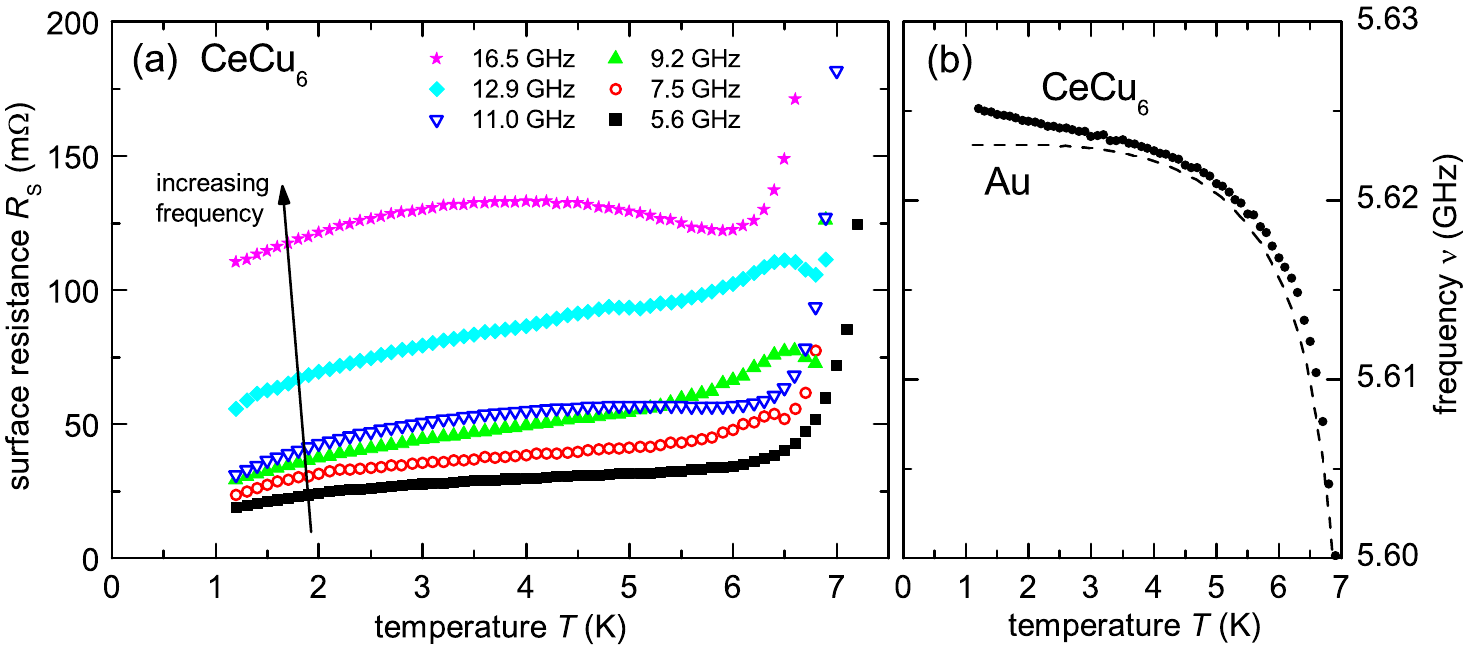}
\end{center}
\caption{(Color online) (a) Temperature dependence of the surface resistance of CeCu$_6$ for different resonator modes. (b) Temperature dependence of the resonance frequency (second mode) for a superconducting lead resonator loaded with the CeCu$_6$ sample compared to a reference measurement with a gold sample. Frequencies obtained for gold were slightly shifted for better comparison.}
\label{f2}
\end{figure}

Nine resonance modes ranging from frequencies of $\nu_{0,1}=1.9\,$GHz to $\nu_{0,9}=16.5$\,GHz could be observed.
The temperature-dependent surface resistance of CeCu$_6$, calculated following Eq.\ (\ref{eq3}), is plotted in Fig.~\ref{f2}(a) for several modes. It has to be noted that above $T\approx6\,$K the increasing lead surface resistance due to thermal population of quasiparticle states close to $T_c$ is no longer negligible, and the values in Fig.~\ref{f2}(a) at these temperatures exceed the actual surface resistance for CeCu$_6$. Upon cooling from 6~K down to 3~K, $R_\textrm{s}$ for all frequencies changes only weakly, with a broad maximum around 5~K and 4~K for the modes at 11 and 16.5~GHz, respectively. Below 3~K, $R_\textrm{s}$ decreases upon cooling for all frequencies, and this decrease becomes more pronounced toward the lowest experimental temperature of 1.2~K. This behavior is similar to that of the dc resistivity of CeCu$_6$, which exhibits characteristic heavy-fermion behavior with a maximum at temperatures around 10~K \cite{Amato1985,Penney1986} and a clear decrease below 3~K.\cite{Amato1985,vLoehneysen1998} The strongest decrease in the dc resistivity, which is a signature of the coherent heavy-fermion state in CeCu$_6$, occurs only at temperatures below 1~K,\cite{Ott1985,Amato1985,Penney1986,vLoehneysen1998} i.e., below our accessible temperature range. However, the downturn in $R_\textrm{s}$ toward the lowest temperatures in Fig.~\ref{f2}(a) does suggest a similar behavior for the microwave response. The observed temperature dependence of the surface resistance demonstrates that our experimental technique is well-suited to address the intrinsic microwave properties of a bulk heavy-fermion metal.

\begin{figure}[tb]
\begin{center}
\includegraphics{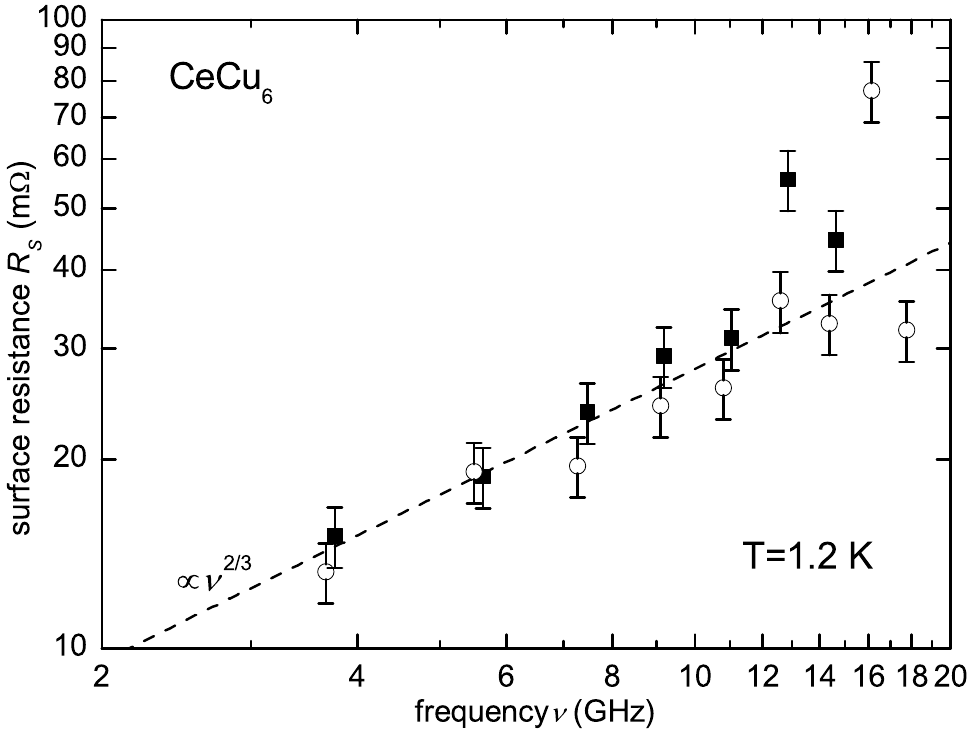}
\end{center}
\caption{Frequency dependence of the surface resistance for a temperature of $T=$1.2\,K as measured with two different resonators (full squares and empty circles). The dashed line is a guideline for a frequency dependence proportional $\nu^{2/3}$, which is expected for a metal in the regime of the anomalous skin effect. The error bars are derived from uncertainties in the $G$ factor.}
\label{f3}
\end{figure}

Fig.~\ref{f3} shows the frequency-dependent surface resistance of CeCu$_6$ at our lowest temperature 1.2~K, as obtained from the data of two different resonators. Up to 12~GHz, $R_\textrm{s}$ can be described by a power-law dependence with exponent 2/3; for higher frequencies the data scattering is larger and does not allow rigorous conclusions whether there are deviations from this power law.
A frequency dependence proportional to $\nu^{2/3}$ is predicted for a good metal which exhibits the anomalous skin effect.\cite{Dressel2002a} In contrast to the normal skin effect, which is governed by local electrodynamics and leads to $R_\textrm{s} \propto \nu^{1/2}$, the anomalous skin effect is a non-local phenomenon.
Here one has to consider the two relevant length scales, namely the mean free path $\ell$ of the conduction electrons and the skin depth $\delta$ (the length scale that characterizes the penetration of the microwave field into the metal).
In the simple local Drude picture one has $\delta \gg \ell$: the scattering sites along the trajectory of an electron are so close in space, at average distance $\ell$, that these frequent scattering events allow the electron to continuously follow the field strength governed by $\delta$. In the opposite regime, $\delta < \ell$, the ballistic motion of an electron between two scattering sites leads to scattering at the latter site that depends on the field distribution at the previous site, i.e., is non-local. In the extreme limit, $\delta \ll \ell$, the response is temperature independent and the surface resistance depends on the Fermi surface.\cite{Ziman1972}

While conventional metals at low temperatures indeed exhibit the anomalous skin effect,\cite{Pippard1947} including the $\nu^{2/3}$ frequency dependence,\cite{DiIorio1988,Hafner2013} this is not expected for typical heavy-fermion metals. Although their relaxation rate is extremely low, in the simple Drude picture (with frequency-independent relaxation rate) this does not translate to a long mean free path: the creation of the heavy-fermion state, with mass enhancement $m^*/m$ and reduction of relaxation rate $\Gamma^*/\Gamma$, is connected to a corresponding reduction of the Fermi velocity.\cite{Scheffler2013,Ebihara1992} As a result, the mean free path between scatterers is not directly affected by the heavy nature of the mobile charges, and at low temperature is of order 100~nm.\cite{Scheffler2005c,Kasahara2005} The low-temperature dc conductivity of a material like CeCu$_6$, of order tens of $\mu\Omega$cm, corresponds to a skin depth of a few $\mu$m at GHz frequencies. Thus one can expect that our CeCu$_6$ sample is in the normal skin-effect regime. This is consistent with the pronounced temperature dependence of $R_\textrm{s}$ down to 1.2~K. 
Furthermore, the resonance frequencies also show a clear temperature dependence, as illustrated in Fig.~\ref{f2}(b) for the second mode. The strong increase below 7~K is due to the temperature-dependent penetration depth of the superconducting lead, which effectively reduces the resonator volume upon cooling. However, below 3~K, the resonator frequency does not change any more due to the superconducting lead, as can be seen from the reference measurement of a Au sample (with anomalous skin effect) with a similar resonator, also shown in Fig.~\ref{f2}(b).\cite{Hafner2013} In contrast, the measurement on CeCu$_6$ displays a pronounced increase of the resonator frequency below 3~K, indicating that the skin depth of the CeCu$_6$ decreases upon cooling, which is inconsistent with being deep in the regime of the anomalous skin effect.

If one can rule out the anomalous skin effect as the origin of $R_\textrm{s} \propto \nu^{2/3}$ in Fig.~\ref{f2}(a), then the explanation has to go beyond the microwave response of conventional metals. The most plausible explanation seems to be a relaxation rate that is frequency dependent even for these low frequencies. Our lowest temperature of 1.2~K corresponds to a photon energy of 25~GHz, therefore it is surprising that the experimental data suggest a frequency-dependent relaxation for frequencies of 5~GHz or even lower.
To reach a full understanding of our data on CeCu$_6$, extensions to both lower and higher temperatures are desired. Our experiment can directly be implemented into a dilution refrigerator, and this should allow access to the well-established Fermi-liquid state of CeCu$_6$,\cite{Ott1985,Amato1985,vLoehneysen1995,vonLoehneysen1996,vLoehneysen1998} where clear theoretical predictions for the frequency-dependent relaxation rate and the consequences for optical properties exist.\cite{Scheffler2013,Gurzhi1959,Maslov2012,Berthod2013} Whether Fermi-liquid properties in $R_s$ at lower temperatures can directly be carried over to our present experiment, at temperatures higher than the observed $T^2$ range in dc resistivity, remains to be seen. Experiments at even higher temperatures, above 6~K, would be helpful to observe the transition between a conventional metal at high temperatures and the correlated state below 10~K. This would require a modification of our setup, going to superconductors with higher $T_c$ than lead, such as Nb or NbN. Microwave measurements along these lines are presently being prepared.

\section{Conclusions}
Our study shows the successful application of simultaneous multi-frequency surface resistance measurements for a CeCu$_6$ bulk single-crystal using superconducting stripline resonators. The measured data for temperatures from 1.2 to 6~K indicate that $R_\textrm{s}$ continuously decreases upon cooling, in consistence with dc resistivity. The observed frequency dependence, with $R_\textrm{s} \propto \nu^{2/3}$, is surprising, as it commonly indicates the anomalous skin effect in good metals at low temperatures, but is inconsistent with the heavy-fermion nature of CeCu$_6$. Therefore, this frequency behavior might indicate a frequency-dependent relaxation rate due to electronic correlations. Future studies both at higher and lower temperatures might solve this open question, and lead to the study of the frequency-dependent microwave response of a Fermi liquid in a well-established model material.

\section*{Acknowledgements}
We thank G.\ Untereiner for the fabrication of the superconducting resonators. This work was supported by the Deutsche Forschungsgemeinschaft (DFG).

%

\end{document}